\documentclass[12pt]{iopart}
\usepackage{graphicx}  

\newcommand{\jpsi}{J/\psi}
\newcommand{\psip}{\psi'}
\newcommand{\ups}{\Upsilon}
\newcommand{\upsp}{\Upsilon'}
\newcommand{\upspp}{\Upsilon''}

\newcommand{\gtrsim}{\geq}
\newcommand{\dNdeta}{\frac{dN_{ch}}{d\eta}\mid_{\eta=0}}
\newcommand{\dNdy}{\frac{dN_{ch}}{dy}\mid_{y=0}}

\begin{document}
\title{High-mass dimuon resonances in Pb-Pb collisions at 
5.5 TeV in CMS}
\author{Dipanwita Dutta for CMS Collaboration}

\address{Nuclear Physics Division, 
Bhabha Atomic Research Centre,
Trombay, Mumbai, India 400085}
\ead{dutta.dipa@gmail.com}
The measurement of the charmonium ($J/\psi$, $\psi'$)
and bottomonium ($\Upsilon$, $\Upsilon'$, ${\Upsilon''}$)
resonances and $Z^0$ boson in nucleus-nucleus collisions
provides crucial information on high density QCD matter.
The observation of anomalous suppression of $J/\psi$
at the CERN-SPS and RHIC is well established but
the clarification of some important questions 
requires equivalent  studies of the $\Upsilon$ family, 
only possible at LHC energies.
The $Z^0$ boson will be produced for the first time
in heavy-ion collisions at the LHC and,
since its dominant production
channel is through $q{\bar q}$
fusion, it is an excellent probe of the
nuclear modification of quark distribution functions.
This paper reports the capabilities of the CMS detector to study 
quarkonium and $Z^0$ production in Pb-Pb 
collisions at 5.5 TeV, through the dimuon decay
channel.


\section{Introduction}

Heavy Ion Collisions at the LHC will explore  regions of energy
and particle density significantly beyond those reachable at
RHIC.
The charmonium  studies in heavy-ion
collisions  at the SPS and at RHIC revealed
 a significant  {\it anomalous} suppression 
 in Pb-Pb and Au-Au collisions at
$\sqrt{s_{NN}}$ = 17.3 GeV~\cite{Abreu:2000} 
and 200 GeV~\cite{Adare:2006}, respectively.
At the LHC energies, the $\Upsilon$ family will become accessible
with much larger cross sections~\cite{cms:ptdr} and
may answer  some unresolved questions on the interpretation of the
quarkonium suppression data.
Heavy quarks are produced
mainly via gluon-gluon fusion, which is sensitive to saturation 
of the gluon density at low-$x$ in the nucleus
(Colour Glass Condensate).
Quarkonium production cross sections in Pb-Pb collisions at the LHC will, thus,
  give information relevant to the study of
the partonic medium and of the initial-state
modifications of the nuclear parton distribution functions.

At the high collision energies of LHC,
the cross section for  processes with $Q^2 > $($50$ GeV)$^2$
is large enough for detailed systematic studies.
 $Z^0$ production proceeds predominantly through the
$q{\bar q}$ channel. Hence, it provides a unique
opportunity to study the modifications of quark distributions in the
nucleus at high $Q^2 = {m^2}_{Z^0}$~\cite{Vogt:2001}.
The CMS detector~\cite{cms:tdr} is designed to
identify and precisely measure muons over a large
energy and rapidity range and is well suited to study
quarkonium and $Z^0$ production in their dimuon decay channel.
 
This work presents the expected capabilities of CMS to measure
the heavy-quarkonia and $Z^0$ cross sections, versus centrality,
rapidity ($y$) and transverse momentum ($p_{\rm T}$), in Pb-Pb collisions
 at $\sqrt{s_{NN}}$ = 5.5~TeV, via their dimuon
decay channel.
The  dimuon mass resolution,
the signal over background ratio and the expected yields 
as a function of $p_{\rm T}$, $y$, and centrality in
one-month Pb-Pb running are presented~\cite{cms:note 2006/089}.
%
\section{Simulation of signal and backgrounds for quarkonia and $Z^0$}
\label{sec:hi_qqbar_simu}
The signal and background dimuons for quarkonium studies
are obtained from realistic distributions:
NLO pQCD for heavy-quark
production processes ~\cite{Bedjidian:2003} and HIJING~\cite{Wang:1991}
 for the soft background. 
The quarkonium production cross sections per 
nucleon-nucleon collision are calculated to NLO 
using the color evaporation model (CEM)~\cite{Bedjidian:2003},
 taking account of shadowing effects, whereas the
$Z^0$ production cross section per nucleon-nucleon collision
 is calculated using PYTHIA6.409~\cite{pythia}.
The Pb-Pb cross sections are obtained by scaling the per
nucleon cross section with $A^2$, where $A=208$.
A main source of background in the dimuon invariant mass spectrum is
combinatorial muon pairs  from the decays of charged
pions and kaons (which
represent about 90\% of the total produced charged particles),
 simulated using input $d^2N/dp_{\rm T}dy$ distributions from HIJING,
for multiplicities $\dNdeta$~=~2500 and 5000
 (lower and upper limit) in the 0--5\% most central Pb-Pb collisions. 
 Another source of background is due to muons from
  open heavy flavour (D,B) meson decays.
The expected average  
multiplicities for  signal high mass dimuon resonances 
 per head-on Pb-Pb collision at $\sqrt s$=5.5 TeV are:
$\jpsi$: 0.034;
$\psip$: 6.2$\times$10$^{-4}$;
$\ups$: 2.1$\times$10$^{-4}$;
$\upsp$: 5.6$\times$10$^{-5}$;
$\upspp$:  3.0$\times$10$^{-5}$; and
$Z^0$: 4.8$\times$10$^{-5}$.
The corresponding values for ``minimum bias collisions'' are: 
6.3$\times$10$^{-3}$;
1.3$\times$10$^{-4}$;
3.8$\times$10$^{-5}$;
1.0$\times$10$^{-5}$;
5.7$\times$10$^{-6}$; and
8.9$\times$10$^{-6}$.
The number of dimuons from D,B meson decays, per head-on Pb-Pb collision,
(with shadowing) is 150 from charm and 5 from beauty.  The values for 
``minimum bias collisions'' are five times smaller.

\subsection{Dimuon Reconstruction and analysis}
\label{sec:hi_qqbar_analysis}
\begin{figure}[htbp]
\centering
\includegraphics[width=0.48\textwidth]{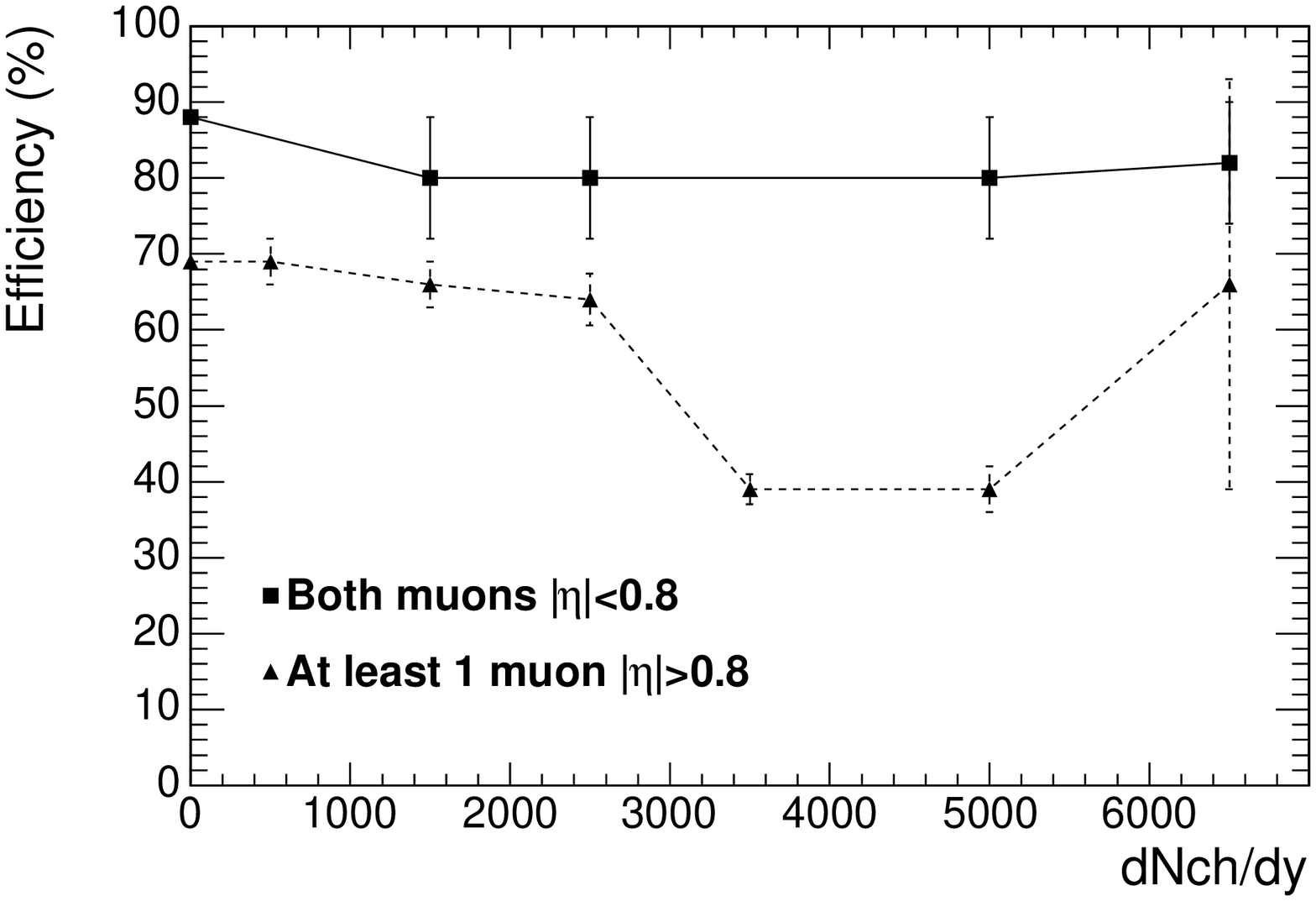}
\includegraphics[width=0.48\textwidth]{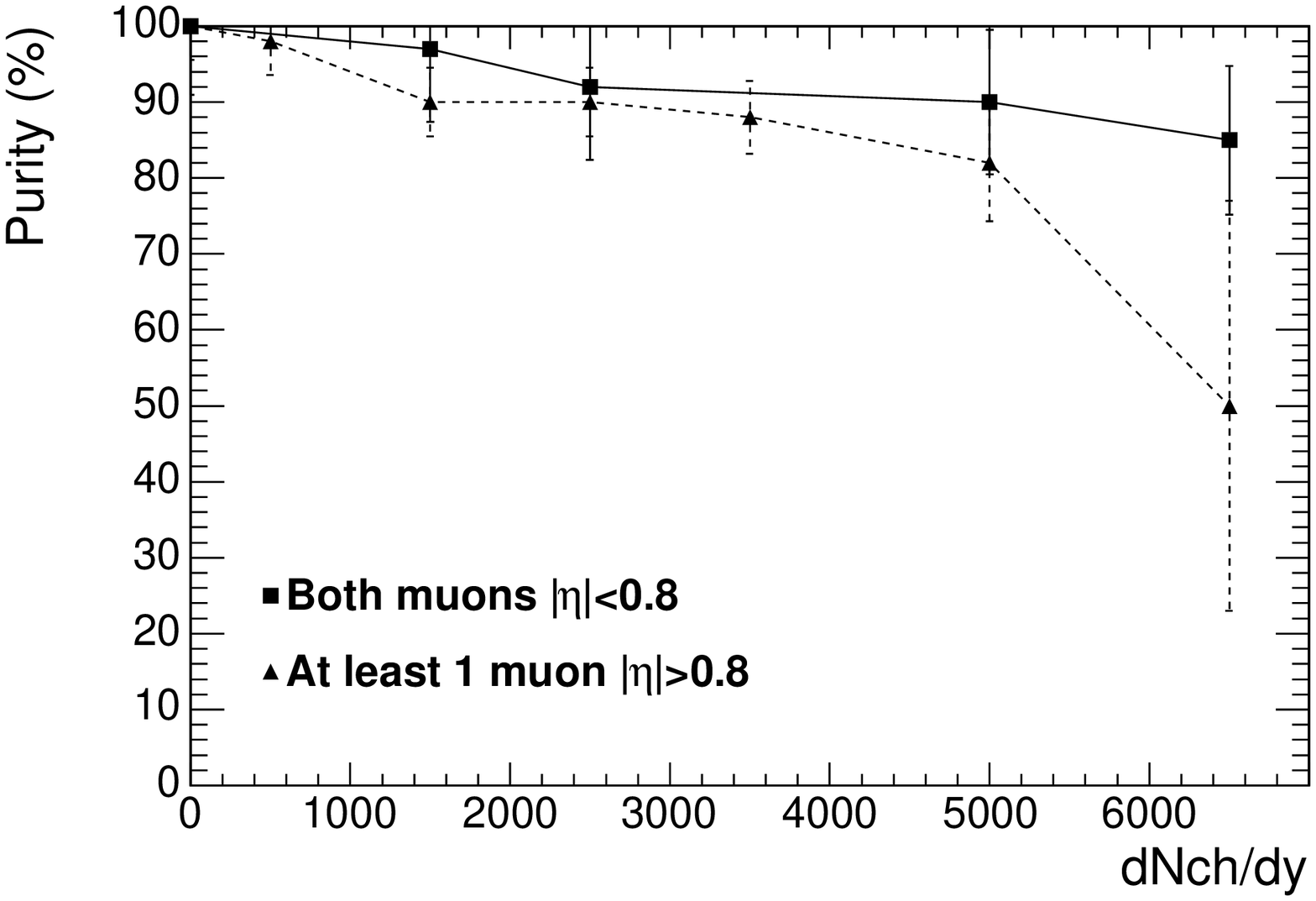}
\caption{$\ups$ reconstruction efficiency (left) and purity (right) as a function of the
Pb-Pb charged particle rapidity density $\dNdy$.}
\label{fig:ups_eff_puri_vs_dNdeta}
\end{figure}

\begin{figure}[htbp]
\centering
\includegraphics[width=0.48\textwidth]{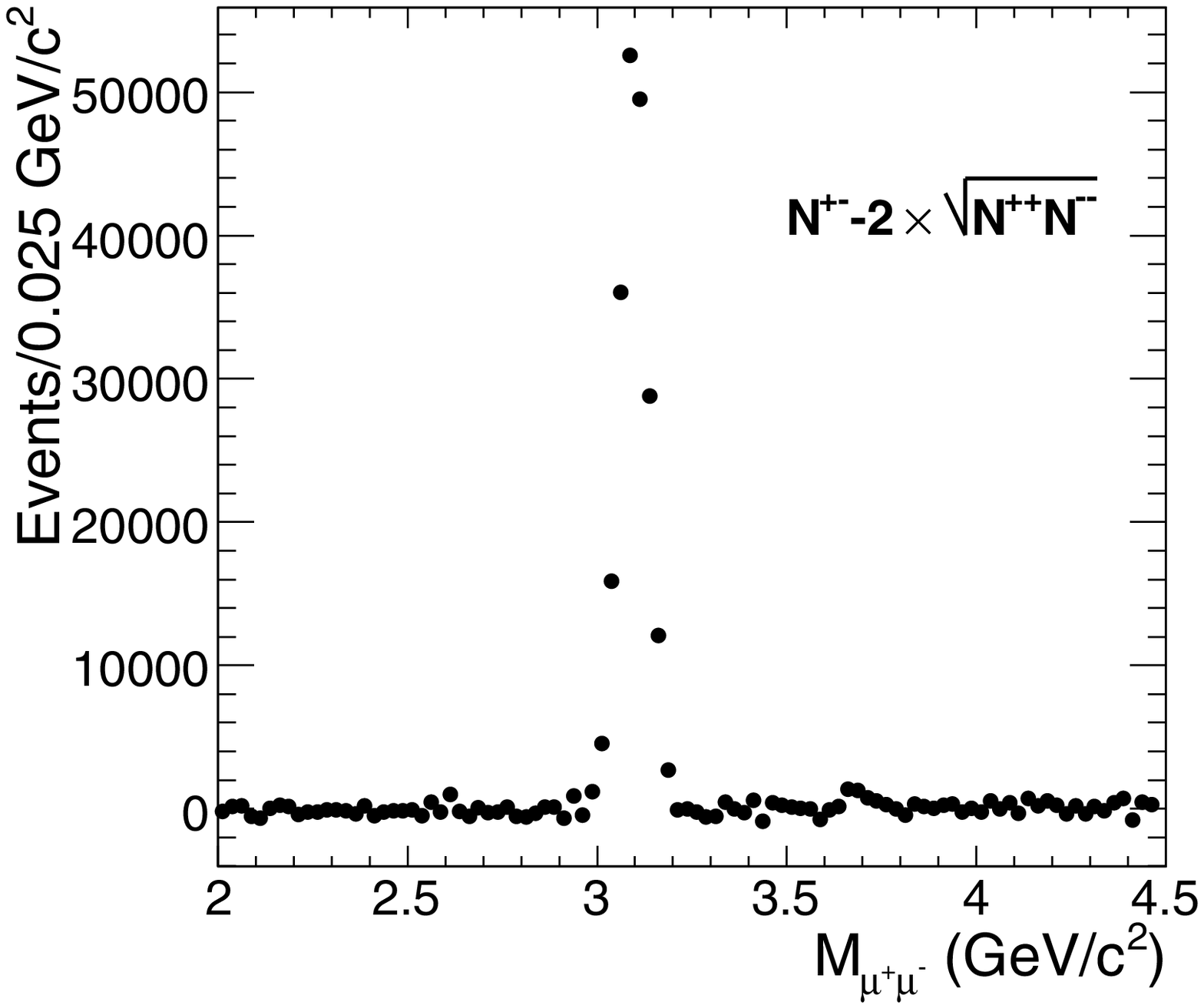}
\includegraphics[width=0.48\textwidth]{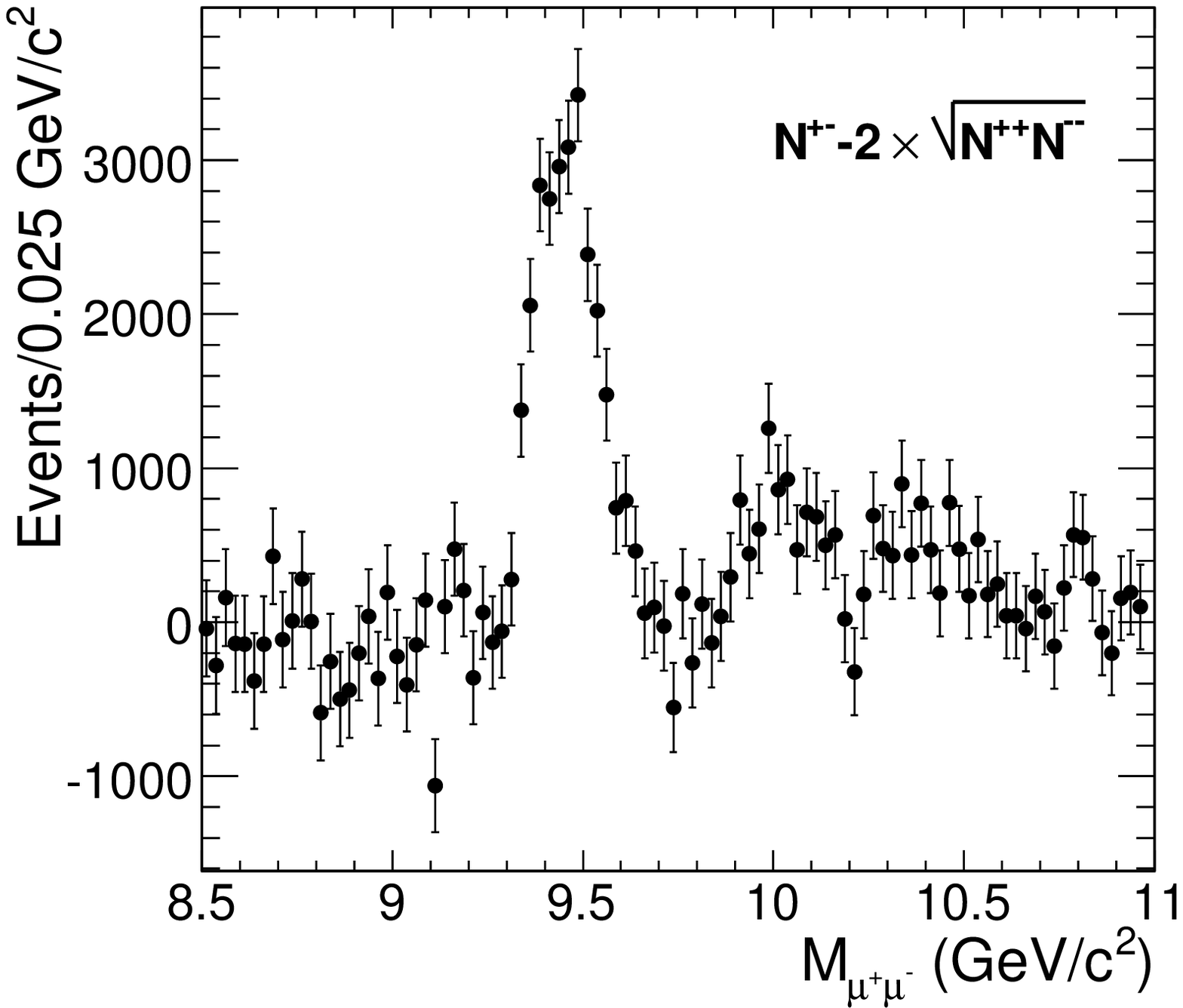}
\caption{Signal dimuon mass distributions in the $\jpsi$ (left)
 and $\ups$ (right) mass regions,
as expected after one month of Pb-Pb running (0.5~nb$^{-1}$)
 for $\dNdeta = 5000$ and
$|\eta|<2.4$, assuming no quarkonium suppression.}
\label{fig:minv_hm_signal_minus_bckgd}
\end{figure}

\noindent
The CMS detector response to muons is parametrised
by 2 dimensional ($p,\eta$) acceptance and trigger tables.
The track is accepted or rejected according to the heavy-ion 
dimuon trigger criteria. The $\jpsi$ and $\ups$ acceptances
were calculated as a function of  $p_T$ for $|\eta|<2.4$. 
The $\jpsi$ acceptance increases with $p_{\rm T}$,
 flattening out at $\sim$~15\% for $p_{\rm T}\gtrsim$ 12 GeV/$c$,
whereas the $\ups$ acceptance starts at $\sim\,$40\% 
 and flattens out at $\sim$~15\% for $p_{\rm T}>4$~GeV/$c$.
The $p_{\rm T}$-integrated acceptance (within $|\eta|<2.4$) is about
1.2\% for the $\jpsi$, 26\% for the $\ups$ and 58\% for the $Z^0$.
The dimuon reconstruction algorithm used in 
the heavy-ion analysis is 
a modification of the regional track finder
 and is based on the muons seeded by the muon stations and on
the knowledge of the primary vertex.
The dependence of the $\ups$ reconstruction efficiency 
on the Pb-Pb charged-particle multiplicity was
obtained from a full GEANT simulation using the
 $\ups$ signal dimuons embedded in HIJING
events. Figure~\ref{fig:ups_eff_puri_vs_dNdeta}
 shows the $\ups$ efficiency
and purity  as a function of the charged-particle density.
In the central barrel, the dimuon reconstruction efficiency remains
above 80\% for all multiplicities. The purity
decreases slightly with increasing $\dNdy$
 but also stays  above 80\% even at multiplicities
as high as $\dNdy$ = 6500. 
About $5\times 10^7$ Pb-Pb collisions were simulated
 with the fast simulation software of CMS, as described in~\cite{cms:ptdr,cms:note 2006/089}.
Muons passing the acceptance tables are combined
to form pairs and each pair is weighted according
to the trigger and reconstruction
efficiencies (dependent on the momentum, pseudorapidity,
purity and event multiplicity), as
determined with the full simulation.
The different quarkonium resonances appear on top of a continuum due to the
random pairing of muons from decays of pions, kaons, charm mesons and bottom mesons.
The background of uncorrelated muon pairs is determined from 
the like-sign muon pairs
mass distribution and subtracted
from the opposite-sign dimuon mass distribution,
giving us a better access to the  quarkonium decay
signals, as shown in Fig.~\ref{fig:minv_hm_signal_minus_bckgd}
for $\dNdeta = 5000$ and $|\eta|<2.4$.
The dimuon mass resolution is about 1\% of the
quarkonium mass: 35~MeV/$c^2$ at the $\jpsi$ mass
 and 86~MeV/$c^2$ at the $\ups$ mass.

\begin{figure}[htbp]
\centering
\includegraphics[width=0.40\textwidth]{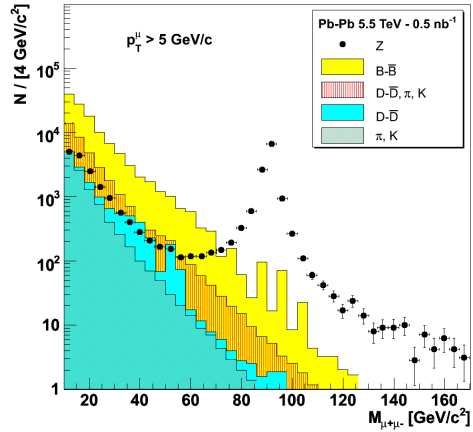}
\caption{Expected $Z^0\rightarrow \mu\mu$ signal and
 combinatorial dimuon background, for
$p_{\rm T}^{\mu} > 5$~GeV/c and $|\eta^{\mu}| < 2.4$, 
 after one month of Pb-Pb running (0.5~nb$^{-1}$ integrated luminosity)}
\label{fig:dimuon_inv}
\end{figure}
Figure ~\ref{fig:dimuon_inv} shows the expected  dimuon
mass distribution from $Z^0$ dimuon decays and from 
several background sources (heavy flavours,
Drell-Yan, pion and kaon decays, and mixed origins).
A clear signal from $Z^0 \to \mu^+\mu^-$ decays is seen, 
of about 11\,000 events within $M_Z \pm 10$ GeV/$c^2$, 
 with less than $5$\% background.

 \section{Summary}
CMS can reconstruct the charmonium and bottomonium
resonances, via their dimuon decay channel,
 with  large acceptances, high efficiencies, good purity and a very
 good dimuon mass resolution, even in the case of exceptionally high
 multiplicities.
The large acceptance in CMS will enable detailed studies
of the nuclear quark distribution functions in a kinematic regime
never probed before.

\end{document}